\documentclass[%
reprint,
superscriptaddress,
amsmath,amssymb,
aps,
pra,
]{revtex4-1}

\usepackage{amsfonts}
\usepackage{bm}
\usepackage{bbm}
\usepackage{mathtools} 
\usepackage{hyperref}

\newcommand{\Htot}{H_{\text{tot}}}
\newcommand{\calA}{\mathcal{A}}
\newcommand{\calJ}{\mathcal{J}}
\newcommand{\calK}{\mathcal{K}}
\newcommand{\calL}{\mathcal{L}}
\newcommand{\calP}{\mathcal{P}}
\newcommand{\calQ}{\mathcal{Q}}
\newcommand{\Rreg}{R_{\text{reg}}}
\newcommand{\ket}[1]{\left| #1 \right\rangle}
\newcommand{\bra}[1]{\left\langle #1 \right|}
\newcommand{\ketbra}[2]{\ket{#1}\!\bra{#2}}

\DeclareMathOperator{\Tr}{Tr}

\newcommand{\stackeq}[1]{\stackrel{\text{\tiny (\ref{#1})}}{=}}

\begin{document}

\title{Time-convolutionless master equation: Perturbative expansions to arbitrary order and application to quantum dots}

\author{Konstantin Nestmann}
    \email{nestmann@physik.rwth-aachen.de}
    \affiliation{Institute of Theoretical Physics, Technische Universit\"at Dresden, 01062 Dresden, Germany}
    \affiliation{Institute for Theory of Statistical Physics, RWTH Aachen, 52056 Aachen, Germany}

\author{Carsten Timm}
    \email{carsten.timm@tu-dresden.de}
    \affiliation{Institute of Theoretical Physics, Technische Universit\"at Dresden, 01062 Dresden, Germany}
    \affiliation{Center for Transport and Devices of Emergent Materials, Technische Universit\"at Dresden, 01062 Dresden, Germany}

\date{\today}

\begin{abstract}
The time-convolutionless quantum master equation is an exact description of the nonequilibrium dynamics of open quantum systems, with the advantage of being local in time. We derive a perturbative expansion to arbitrary order in the system-reservoir coupling for its generator, which contains significantly fewer terms than the ordered-cumulant expansion. We show that the derived expansion also admits a simple recursive formulation. The derived series is then used to describe the nonequilibrium dynamics of a quantum dot, including coherences. We find a relation between the generator and a generalization of the \textit{T}-matrix. Even though the \textit{T}-matrix rate equations are plagued by divergences, we show that these cancel order by order in the generator of the time-convolutionless master equation. This generalizes previous work on the time-convolutionless Pauli master equation, which does not include coherences, and lays the ground for possible numerical evaluations and analytical resummations.
\end{abstract}

\maketitle

\section{Introduction}

The description of open quantum systems out of equilibrium is of fundamental interest within theoretical physics. One aim is to understand how a quantum system in contact with several reservoirs evolves into a stationary nonequilibrium state and what the properties of this state are. In particular, for small systems such as quantum dots, the stationary state can be very different from any equilibrium state. As artifical atoms, quantum dots are experimentally controllable systems and therefore allow the study of phenomena that play a fundamental role in nanoelectronics, spintronics, dissipative quantum systems, and quantum information processing~\cite{HansonEtAl07, AndergassenEtAl10}.

In a situation where the small system (e.g., the quantum dot) can be described by a limited number of degrees of freedom, which may be strongly interacting, and where the system-reservoir coupling (e.g., the hybridization) is weak, it is reasonable to focus on the dynamics of the reduced density matrix $\rho_S(t)$. By integrating out the reservoirs states one naturally obtains a perturbative series in the system-reservoir coupling, while the interactions within the small system can, in principle, be treated exactly. A differential equation that describes the evolution of the reduced density matrix is called a \textit{quantum master equation} (ME)~\cite{BreuerPetruccione02}.

A ME can either be nonlocal in time, which means that the rate of change of the reduced density matrix at a time $t$ depends on all the states of the reduced density matrix prior to $t$, or it can be local in time (``time convolutionless,'' TCL) \cite{Shibata77, Chaturvedi79, Shibata80}, such that the rate of change of the reduced density matrix at time $t$ depends only on the reduced density matrix at time $t$ itself. A time-nonlocal ME, for example of Nakajima-Zwanzig (NZ) type \cite{Nakajima58, Zwanzig60, Zwanzig64}, has the general form
\begin{equation}
\dot\rho_S(t) = \int_{t_0}^t ds\, K(t,s)\, \rho_S(s) ,
\label{eq:time_nonlocal_me}
\end{equation}
where the central object is the so-called memory kernel $K(t,s)$. The memory kernel is a superoperator, which acts on the reduced density matrix at past times $s$. Given a method to generate all terms in a perturbation series for the memory kernel, one may try to resum a \mbox{(sub-)}series. Using a real-time diagrammatic scheme introduced by Schoeller, Sch\"on, and K\"onig \cite{Schoeller94, Koenig95, Koenig96}, all terms of the memory kernel can be generated. Furthermore, Schoeller \cite{Schoeller09} presented an approach in Laplace space that is suitable for a nonequilibrium renormalization-group treatment, and, in principle, includes all orders in the system-reservoir coupling.

The clear advantage of the TCL ME of the general form $\dot\rho_S(t) = S(t,t_0)\, \rho_S(t)$ with the generator $S(t,t_0)$ is that it is structurally much simpler then a ME of NZ type, while still being exact and describing the full dynamics \cite{BreuerPetruccione02}. The central challenge in the TCL approach lies in the expansion of the generator of the TCL ME in the system-reservoir coupling. In addition to an expansion using ordered cumulants introduced by van Kampen \cite{vanKampen74}, which holds for arbitrary Hamiltonians and time dependences but is inconvenient for practical calculations due to the enormous growth of the number of terms for higher orders, one of us \cite{Timm11} has considered a Pauli TCL ME that describes the \textit{diagonal} entries of the reduced density matrix. In Ref.\ \cite{Timm11}, a scheme to construct all orders in the perturbative expansion of the Pauli TCL generator was derived. While this scheme does not assume vanishing coherences (off-diagonal terms of the reduced density matrix), it eliminates them from the description and thus does not provide any information on them. Even though a close connection to the \textit{T}-matrix rate equations, which are plagued by divergences, was found, it was shown that all divergences in the Pauli TCL generator cancel each other order by order~\cite{Timm11}.

In this paper, we first derive a new perturbative expansion for the generator of the full TCL ME with coherences. This expansion admits a simple recursive formulation and is equivalent to ordered cumulants \cite{vanKampen74} but can be thought of as a reorganization of the terms that leads to dramatic simplifications. Therefore, it has the same range of validity as ordered cumulants and holds for arbitrary Hamiltonians and time dependences. Using the derived series, we generalize the approach of Ref.\ \cite{Timm11} to describe the dynamics of a typical quantum-dot system including coherences. As in the Pauli case, we show that each order of the TCL generator consists of several diverging parts, all of which cancel, leaving finite corrections such that every order of the TCL generator is well defined.

\section{The TCL master equation}

In order to introduce our notation and make contact with the superoperator formalism, we briefly review the derivation of the TCL ME using projection operators \cite{Shibata77, Chaturvedi79, Shibata80, BreuerPetruccione02, Timm08, Timm11}. We then derive simpler expansion formulas for the generator of this ME.

\subsection{Derivation}

We consider a large system consisting of a small subsystem $S$ of interest and an environment or bath $B$. The full Hamiltonian reads as
\begin{equation}
\Htot(t) = H_S \otimes \mathbbm{1}_B + \mathbbm{1}_S \otimes H_B + H_V(t),
\end{equation}
where we assume for simplicity that the Hamiltonians of the small system and of the bath are time independent. In the following, we will suppress identity operators of the system, $\mathbbm{1}_S$, and of the bath, $\mathbbm{1}_B$. Then, the decoupled system is described by $H_0 \coloneqq H_S+H_B$. The evolution of the full system is governed by the von Neumann equation. Defining the Liouvillian superoperators $\calL_X(t) \, \bullet \coloneqq [H_X(t), \bullet]$, where $X \in \{ S, B, 0, V \}$, the von Neumann equation in the Schr\"odinger picture can be written as
\begin{equation}
i \dot \rho^{(S)}(t) = [ \calL_0 + \calL_V(t) ]\, \rho^{(S)}(t).
\end{equation}
Here, the superscript ``$(S)$'' indicates the Schr\"odinger picture. It will, however, be more convenient to work in the interaction picture, where the von Neumann equation takes the form 
\begin{equation}
i \dot \rho(t) = e^{i \calL_0 t} \calL_V(t) e^{-i \calL_0 t}\, \rho(t) \eqqcolon \calL(t)\, \rho(t).
\end{equation}
The unitary evolution of the full density matrix can formally be written as
\begin{equation}
\rho(t) = G(t,t_0)\, \rho(t_0),
\label{eq:G.0}
\end{equation}
with the full propagator $G$, which is the superoperator solution of the differential equation $\partial_t G(t, t_0) = -i \calL(t)\, G(t,t_0)$ with the initial condition $G(t_0, t_0) = \mathcal{I}$. It is explicitly given by
\begin{equation}
G(t, t_0) = T_{\leftarrow} \exp \left(-i \int_{t_0}^t ds\ \calL(s) \right),
\label{eq:G.1}
\end{equation}
where $T_\leftarrow$ is the time-ordering directive, which orders all following factors such that their time arguments increase from right to left. $G(t,t_0)$ is the superoperator equivalent of the time-evolution operator for state vectors.

Since we are only interested in observables describing the relevant (small) system $S$, it is sufficient to investigate the reduced density matrix 
\begin{equation}
\rho_S(t) \coloneqq \Tr_B \rho(t) .
\end{equation}
We proceed by introducing a projection superoperator $\calP$ onto the subspace of the relevant degrees of freedom,
\begin{equation}
\calP \, \bullet \coloneqq \Tr_B( \bullet ) \otimes \rho_B,
\end{equation}
where $\rho_B$ is an arbitrary reference density matrix of the bath. If the initial full state is not entangled, i.e., the initial state is a product state $\rho(t_0)=\rho_S^0 \otimes \rho_B^0$, it is convenient to choose $\rho_B=\rho_B^0$, and we will implicitly do so. Furthermore, we denote the projection onto the irrelevant degrees of freedom by $\calQ \coloneqq \mathcal{I}-\calP$. It is easy to check that $\calP$ and $\calQ$ are indeed orthogonal projection superoperators satisfying $\calP^2=\calP$, $\calQ^2=\calQ$, and $\calP \calQ = \calQ \calP = 0$.  Moreover, the following properties hold:
\begin{eqnarray}
\calP \calL_B &=& \calL_B \calP = 0, \\
\left[ \calP, \calL_0 \right] &=& \left[ \calP, \calL_S \right] = 0.
\end{eqnarray}
With the help of the projection $\calQ$, we here define the irreducible propagator $G_{\calQ}$ as the solution of the differential equation $\partial_t G_{\calQ}(t,t_0)=-i\calQ \calL(t)\, G_{\calQ}(t,t_0)$ with the initial condition $G_{\calQ}(t_0,t_0)=\mathcal{I}$. The solution can be written as
\begin{equation}
G_{\calQ}(t,t_0) = T_{\leftarrow} \exp \left(-i
        \calQ \int_{t_0}^t ds\, \calL(s) \right).
\end{equation}

In order to derive the TCL ME, the von Neumann equation is split into two parts:
\begin{eqnarray}
\frac{\partial}{\partial t}\, \calP \rho(t)
    &=& -i \calP \calL(t)\, \calP \rho(t)
       -i \calP \calL(t)\, \calQ \rho(t), \label{eq:split_von_neumann_1} \\
\frac{\partial}{\partial t}\, \calQ \rho(t)
    &=& -i \calQ \calL(t)\, \calP \rho(t)
       -i \calQ \calL(t)\, \calQ \rho(t).
\end{eqnarray}
From now on, we assume that the initial state is a product state, which implies that $\calQ \rho(t_0)=0$. Then the second equation is solved by
\begin{equation}
\calQ \rho(t) = -i \int_{t_0}^t ds\, G_{\calQ}(t,s) \calQ\calL(s)\, \calP\rho(s). \label{eq:q_rho_nonlocal}
\end{equation}
Now the main idea lies in writing the density matrix at times $s\leq t$ as the density matrix at time $t$ propagated backwards in time, that is
\begin{equation}
\rho(s)=G^{-1}(t,s)\, \rho(t) = G^{-1}(t,s)\, ( \calP + \calQ ) \rho(t), \label{eq:deriv_q_rho_local}
\end{equation}
and again solving Eqs.~(\ref{eq:q_rho_nonlocal}) and (\ref{eq:deriv_q_rho_local}) together for the irrelevant part, which gives
\begin{equation}
\calQ \rho(t) = \left[ \mathcal{I} - \Sigma(t,t_0) \right]^{-1} \Sigma(t,t_0)\, \calP \rho(t),
\label{eq:q_rho_local}
\end{equation}
where we have defined the auxiliary superoperator
\begin{equation}
\Sigma(t,t_0) \coloneqq -i \int_{t_0}^t ds\, G_{\calQ}(t,s) \calQ\calL(s)\calP G^{-1}(t, s) .
\label{eq:def_sigma_integral}
\end{equation}
Note that there is no relation between $\Sigma$ and the memory kernel, which we denote as $\Sigma_K$ in this paper, see Eq.~(\ref{eq:time_nonlocal_me}). Inserting Eq.~(\ref{eq:q_rho_local}) into Eq.~(\ref{eq:split_von_neumann_1}) results in the TCL ME
\begin{equation}
\frac{\partial}{\partial t}\, \calP \rho(t) = S(t, t_0)\, \calP \rho(t), 
\end{equation}
where the TCL generator $S$ is given by
\begin{equation}
S(t,t_0) \coloneqq -i \calP \calL(t) \left[ \, \mathcal{I} - \Sigma(t,t_0)\right]^{-1} \calP.
\label{eq:tcl_generator_classic_result}
\end{equation}
Since no approximations were used in the derivation, the TCL ME is exact. However, it relies on the existence of the inverse of $\mathcal{I}-\Sigma$. Since $\Sigma(t_0,t_0)$ vanishes and the matrix elements of $\Sigma$ are continous functions of time, the existence of the inverse is at least guaranteed for sufficiently small $t-t_0$~\cite{BreuerPetruccione02}.

Since usually there does not exist a closed form for the TCL generator \cite{Reimer19}, a perturbative approach is useful. It can be shown that the general $n$th-order contribution $S^{(n)}$ in the system-reservoir coupling of the TCL generator can be expressed using so-called ordered cumulants \cite{vanKampen74, BreuerPetruccione02}, which were originally introduced by van Kampen \cite{vanKampen74} for the description of stochastic differential equations. One finds that the $n$th order of $S$ is given by
\begin{eqnarray}
S^{(n)}(t,t_0) &=& (-i)^n \int_{t_0}^t dt_1 \cdots \int_{t_0}^{t_{n-2}} dt_{n-1} \nonumber \\
&&{}\times \big\langle \calP \calL(t) \calL(t_1) \cdots \calL(t_{n-1}) \calP \big\rangle_\mathrm{oc},
\label{eq:expansion_tcl_ordered_cumulants}
\end{eqnarray}
with the ordered cumulants 
\begin{eqnarray}
\lefteqn{\big\langle \calP \calL(t) \calL(t_1) \cdots \calL(t_{n-1})\calP \big\rangle_\mathrm{oc} \coloneqq \sum_{\mathrm{oc}\:q} (-1)^q\, \calP \calL(t) \cdots} \nonumber \\
&&\quad{}\times \calL(t_i) \calP \cdots \calP \calL(t_j) \cdots \calL(t_k) \calP.\hspace{7em}
\label{eq:def_ordered_cumulants}
\end{eqnarray}
They are constructed according to the following rules: first, one writes down a string of $n$ Liouvillians $\calL$ (with projections $\calP$ before and after) and inserts $q\le n-1$ projections $\calP$ between them. Then, the Liouvillians are furnished with time arguments: the first Liouvillian from the left always gets the argument $t$, while the others carry every permutation of $t_1, \dots, t_{n-1}$, with the restriction that between two successive $\calP$ insertions the time arguments are ordered chronologically decreasing from left to right. Therefore, in Eq.~(\ref{eq:def_ordered_cumulants}) we have $t>\cdots>t_i$ and $t_j>\cdots>t_k$. Finally, $\sum_{\mathrm{oc}\:q}$ sums over all possible terms that follow these rules.

Since we later want to compare the ordered-cumulant expansion to other expansions, it is of interest to ask how many terms one must add to obtain $S^{(n)}$. We focus on the general case in which this number is not reduced by symmetries. Then, a term with $n$ Liouvillians $\calL$ and $q$ projections $\calP$ can be thought of as consisting of $q+1$ distinct sections. Given such a term with the $j$th section consisting of $n_j$ Liouvillians there are
\begin{equation}
\binom{n-1}{n_1, n_2, \dots, n_{q+1}} = \frac{(n-1)!}{n_1! n_2 !\cdots n_{q+1}!}
\end{equation}
possibilities to distribute the time arguments. Thus, to obtain $S^{(n)}$, one needs to add 
\begin{equation}
\sum_{q=0}^{n-1}\;
\sum_{\substack{
        n_1+n_2+\cdots+n_{q+1} = n-1 \\
        n_1 \geq 0; \ n_2,\dots,n_{q+1}\geq1 }}
    \binom{n-1}{n_1, n_2, \dots, n_{q+1}}
\end{equation}
terms. For $n=4$ this evaluates to $26$ terms, for $n=6$ to $1082$, and for $n=10$ to over 14 million. It can be shown that, for large $n$, the number of terms asymptotically approaches~\cite{OEISA000629}
\begin{equation}
\frac{(n-1)!}{(\ln 2)^n}
\sim \sqrt{\frac{2 \pi e}{\ln 2}} \left( \frac{n-1}{e \ln 2} \right)^{n-1/2}
\end{equation}
This rapid growth makes the expansion in ordered cumulants impractical for numerical calculations or analytical partial resummations.

\subsection{Simplified forms of the TCL generator}

The integral in Eq.~(\ref{eq:def_sigma_integral}), which we used to define the superoperator $\Sigma$, can be solved explicitly, as was shown in Ref.~\cite{Timm08}. However, for the TCL generator itself, further simplifications are possible, which were not exploited yet and which we will present in this section.

For completeness, we first repeat the steps taken in Ref.~\cite{Timm08}. To do so, we need the derivatives of $G_{\calQ}$ and $G^{-1}$ with respect to their second time arguments:
\begin{eqnarray}
\frac{\partial}{\partial s}\, G_{\calQ}(t,s) &=& i G_{\calQ}(t, s) \calQ \calL(s),
\label{eq:deriv_propagator_lower_GQ}
\\
\frac{\partial}{\partial s}\, G^{-1}(t,s) &=& -i \calL(s) G^{-1}(t,s).
\label{eq:deriv_propagator_lower_Gi}
\end{eqnarray}
Now we can write the integrand of Eq.~(\ref{eq:def_sigma_integral}) as
\begin{eqnarray}
\lefteqn{ -i G_{\calQ}(t,s) \calQ\calL(s)\calP G^{-1}(t, s) } \nonumber \\
&&\quad = G_{\calQ}(t,s) \calQ\, \frac{\partial}{\partial s}\, G^{-1}(t, s) + \left[ \frac{\partial}{\partial s}\, G_{\calQ}(t,s) \right]\! \calQ G^{-1}(t, s) \nonumber \\
&&\quad = \frac{\partial}{\partial s}\, G_{\calQ}(t,s) \calQ G^{-1}(t, s).
\end{eqnarray}
Thus, $\Sigma$ is given by 
\begin{equation}
\Sigma(t,t_0) = \calQ - G_{\calQ}(t,t_0) \calQ G^{-1}(t, t_0).
\label{eq:deriv_tcl_simple_1}
\end{equation}

We now turn to the simplification of the TCL generator itself. Inserting Eq.~(\ref{eq:deriv_tcl_simple_1}) into Eq.~(\ref{eq:tcl_generator_classic_result}) leads to
\begin{eqnarray}
S &=& -i \calP \calL \left[ \calP + G_{\calQ} \calQ G^{-1} \right]^{-1} \calP \nonumber \\
&=& -i \calP \calL \left[ \left( \calP G + G_{\calQ} \calQ \right) G^{-1} \right]^{-1} \calP \nonumber \\
&=& \calP \dot G \left[ \calP G + G_{\calQ} \calQ \right]^{-1} \calP,
\label{eq:deriv_tcl_simple_2}
\end{eqnarray}
where we evaluate all superoperators at $t$ and $t_0$ and the overdot denotes the derivative with respect to the first time argument. Now, the crucial step lies in the realization that
\begin{equation}
\left[ \calP G + G_{\calQ} \calQ \right]^{-1} \calP = \calP \left[ \calQ + \calP G \calP \right]^{-1},
\label{eq:simplification_inverse}
\end{equation}
which we prove in Appendix \ref{app:proof_identity_inverse}. The term $G_{\calQ} \calQ$ inside the inverse therefore does not contribute to $S$ and we arrive at the result that the TCL generator is given by
\begin{equation}
S(t,t_0) = \calP \dot G(t,t_0) \calP \left[ \calQ + \calP G(t,t_0) \calP \right]^{-1}.
\end{equation}
Remarkably, it is therefore possible to express the TCL generator through the (reduced) propagator $\calP G \calP$ alone, which is much simpler then expressing it through the inverse of a complicated integral such as Eq.~(\ref{eq:def_sigma_integral}). By defining
\begin{equation}
\Pi(t,t_0) \coloneqq \calQ + \calP G(t,t_0) \calP,
\end{equation}
an even more compact rewriting is possible:
\begin{equation}
S(t,t_0) = \dot \Pi(t,t_0)\, \Pi^{-1}(t,t_0).
\end{equation}
This suggests to interpret $\Pi$ as the propagator of the open system, which makes the physical content of the TCL generator obvious: when the generator $S$ acts on the state $\calP \rho(t)$, this state is first propagated backwards in time to reconstruct its initial state $\calP \rho(t_0)$. It is then propagated forward in time back to $t$ and the derivative is taken. All the memory effects enter into the TCL generator through this double propagation, which allows us to formulate an exact time-local master equation even in non-Markovian situations. Furthermore, we now see that the existence of the complicated inverse $\mathcal{I}-\Sigma$ is equivalent to the existence of the inverse propagator~$\Pi^{-1}$.

The previous considerations enable us to derive a very efficient series expansion in orders of the coupling Liouvillian. The propagator can be expanded as a Dyson series as $G(t,t_0) = \mathcal{I} + \bar G(t,t_0)$ with
\begin{equation}
\bar G(t,t_0) \coloneqq -i \int_{t_0}^{t} ds\, \calL(s) G(s, t_0) = \sum_{n=1}^\infty G^{(n)}(t,t_0)
\end{equation}
and
\begin{equation}
G^{(n)}(t,t_0) = (-i)^n \int_{t_0}^{t} dt_1 \cdots \int_{t_0}^{t_{n-1}} dt_n\, \calL(t_1) \cdots \calL(t_n).
\label{eq:orders_propagator}
\end{equation}
Therefore, the TCL generator can be written as
\begin{eqnarray}
S &=& \calP \dot{\bar G} \calP \left[ \mathcal{I} + \calP \bar G \calP \right]^{-1} \nonumber \\
&=& \sum_{n=0}^\infty (-1)^n
    \mkern-12mu \sum_{\mu_0, \dots, \mu_n = 1}^\infty \mkern-12mu 
    \calP \dot G^{(\mu_0)} \calP G^{(\mu_1)} \calP \cdots \calP G^{(\mu_n)} \calP .\hspace{1.5em}
\label{eq:orders_tcl_explicit}
\end{eqnarray}
It follows that the $n$th order of the TCL generator is given by 
\begin{equation}
S^{(n)} = \sum_{j=0}^{n-1} (-1)^{j}
    \mkern-12mu  \sum_{\mu_0 + \dots + \mu_j = n} \mkern-12mu 
    \calP \dot G^{(\mu_0)} \calP G^{(\mu_1)} \calP \cdots \calP G^{(\mu_j)} \calP.
\label{eq:orders_tcl_explicit_n}
\end{equation}
Moreover, the orders $S^{(n)}$ of Eq.~(\ref{eq:orders_tcl_explicit_n}) satisfy the remarkable recursion relation
\begin{equation}
S^{(n)} = \calP \dot G^{(n)} \calP - \sum_{j=1}^{n-1} S^{(n-j)} \calP G^{(j)} \calP,
\label{eq:orders_tcl_recursive}
\end{equation}
with the initial condition $S^{(1)}=\calP \dot G^{(1)} \calP=-i\calP \calL \calP$. The proof is relegated to Appendix \ref{app:proof_recursive_expansion}. Note that the expansions in Eqs.~(\ref{eq:orders_tcl_explicit_n}) and (\ref{eq:orders_tcl_recursive}) must be equivalent, order by order, to the expansion using ordered cumulants in Eq.~(\ref{eq:expansion_tcl_ordered_cumulants}) because they all expand in the same parameter. The number of terms in Eq.~(\ref{eq:orders_tcl_explicit_n}) is given by the number of compositions of the number $n$, which is equal to $2^{n-1}$, which is some improvement compared to the series of ordered cumulants, where the number of terms of order $n$ scales like $(n-1)^{n-1/2}$ for large $n$. A more remarkable simplification is realized by the recursion relation (\ref{eq:orders_tcl_recursive}) since only $n$ terms contribute at the $n$th order.

\section{Nonequilibrium dynamics of a quantum dot}

In this section, we turn to the description of a quantum dot coupled to leads under a bias voltage, which keeps the system out of equilibrium. After introducing the model, we use the results of the previous section to construct a series expansion of the generator and then show that it is finite order by order.

\subsection{Model}

We consider a quantum dot coupled to leads in such a way that the hybridization between them is adiabatically turned on. We model the system with a Hamiltonian of the form
\begin{equation}
\Htot(t) = H_S + H_B + e^{\eta t} H_V \eqqcolon H_0 + e^{\eta t} H_V.
\label{eq:total_hamiltonian}
\end{equation}
Here, $H_S$ describes the isolated dot. We do not make assumptions about possible degeneracies in the dot's energy spectrum. It is attached to electrodes, which we take as free-electron reservoirs:
\begin{equation}
H_B = \sum_{r, \sigma, k} \epsilon_{r \sigma k}\, a_{r \sigma k}^{\dag} a_{r \sigma k}.
\end{equation}
The operator $a_{r \sigma k}^{\dag}$ creates an electron in the reservoir $r \in \{\text{left}, \text{right}\}$ with spin $\sigma$ and orbital index $k$, which contains all other quantum numbers. The junctions connecting the dot to the reservoirs are modeled by the hybridization Hamiltonian
\begin{equation}
H_V = \sum_{r, \sigma, k, \nu} t_{r \sigma k \nu}\, a_{r \sigma k}^{\dag} c_{\sigma \nu} + \text{H.c.},
\label{eq:hamiltonian_v}
\end{equation}
where $c^{\dag}_{\sigma \nu}$ creates an electron in molecular orbital $\nu$ with spin $\sigma$. The hybridization carries the time-dependent prefactor $e^{\eta t}$ [see Eq.\ (\ref{eq:total_hamiltonian})]. We take $t_0\rightarrow -\infty$ and subsequently $\eta \rightarrow 0^{+}$, which corresponds to adiabatically switching on the coupling between dot and leads. Below, we will see that the prefactor $e^{\eta t}$ allows us to count diverging contributions in powers of $\eta^{-1}$ and thus to work with well-defined intermediate expressions at finite $\eta$. As before, we take the initial state to be a product state. Note that $\calP \calL_V \calP = 0$ holds because the expression inside the trace introduced by the leftmost $\calP$ is guaranteed to only contain off-diagonal elements. Physically, this trace is the average of a single fermionic bath operator $a_{r\sigma k}$ or $a_{r\sigma k}^\dag$ in the bath reference state, which vanishes.

\subsection{Perturbative expansion in the hybridization}

The goal is now to derive the TCL generator $S$. Then, the nonequilibrium dynamics of the quantum dot follows from the TCL ME and the stationary state could in principle be obtained as the right eigenvector of $S$ to the eigenvalue zero. As we have remarked earlier, the existence of the TCL generator is formally only certain for sufficiently small $t-t_0$. We are thus investigating a somewhat extreme situation, as in our case $t-t_0 \rightarrow \infty$ and there is no guarantee that the generator exists.

We start from Eq.~(\ref{eq:orders_tcl_explicit}). First, we need to construct the expansion of $\calP \dot G \calP$. For convenience, we define
\begin{equation}
R(t,t_0) \coloneqq \calP \dot G(t,t_0) \calP = -i \calP \calL(t) G(t,t_0) \calP.
\label{eq:r_definition}
\end{equation}
This quantity can be interpreted as a generalized \textit{T}-matrix generator for the following reason: if one considers the equation of motion $\partial_t \calP \rho(t) = R(t,t_0)\, \calP \rho(t)$, which of course does not give the correct time evolution \cite{footnote_RE}, and then only keeps the diagonal elements, one obtains the usual \textit{T}-matrix rate equations~\cite{Timm11}.

Because of the simple exponential time dependence in the Liouvillian, it is straightforward to integrate Eq.~(\ref{eq:orders_propagator}), which for $t_0\to-\infty$ yields
\begin{eqnarray}
\lefteqn{R^{(\mu)} \ketbra{p}{q} = -i\, e^{\mu \eta t} e^{i \calL_S t}\, \calP \calL_V\, \frac{1}{\Delta E_{p'q'} - \calL_0 + (\mu-1) i \eta}} \nonumber \\
&&{}\times \calL_V \cdots \calL_V\, \frac{1}{\Delta E_{p'q'} - \calL_0 + i \eta}\, \calL_V \calP e^{-i \calL_S t} \calP \ketbra{p}{q}.\hspace{2em}
\label{eq:orders_r}
\end{eqnarray}
Here, we use the convention that $\ket{p}=\ket{p'} \otimes \ket{p''}$, where primed states $\ket{p'}$ belong to the quantum-dot Hilbert space, while double-primed states $\ket{p''}$ belong to the lead Hilbert space. In the same spirit, $\Delta E_{p'q'}$ stands for the energy difference of the dot states $\ket{p'}$ and $\ket{q'}$.

Since the leftmost $\calP$ in Eq.~(\ref{eq:orders_r}) traces over lead states and the rightmost $\calP$ contains the equilibrium initial lead density matrix $\rho_B^0$, $R^{(\mu)}$ contains equilibrium averages of products of lead creation (annihilation) operators $a^{\dag}_{r \sigma k}$ ($a_{r \sigma k}$). To obtain nonzero contributions, these operators must be paired. Thus, all odd orders $R^{(2\mu+1)}$ vanish.

For Eq.\ (\ref{eq:orders_tcl_explicit_n}) we also need the expansion of $\calP G \calP$. Using a superoperator index notation with $\calA_{mn}^{pq} = \bra{m} \left( \calA \ket{p}\!\bra{q} \right) \ket{n}$, the orders of $\calP G \calP$ can be expressed through Eq.~(\ref{eq:orders_r}):
\begin{equation}
\big( \calP G^{(\mu)} \calP \big)_{mn}^{pq} = \frac{1}{i \left( \Delta E_{m' n'} - \Delta E_{p' q'} \right) + \mu \eta}\, \big( R^{(\mu)} \big)_{mn}^{pq}.
\label{eq:connection_r_pgp_coordinates}
\end{equation}
A divergence evidently occurs in the limit $\eta \rightarrow 0^+$ if $\Delta E_{m' n'} = \Delta E_{p' q'}$. This happens if the superoperator maps a coherence with a certain oscillation frequency $\Delta E_{p'q'}/h$ (possibly zero) onto one with the same oscillation frequency or if it maps a diagonal contribution ($p'=q'$) onto a diagonal contribution ($m'=n'$). The latter case also occurs for the Pauli ME \cite{Timm11}. The form of the $\eta$ dependence turns out to be inconvenient for the analysis of the limit. To extract the divergent part and obtain pure negative powers of $\eta$, we define superoperators $\calJ_\mu$ through their matrix elements:
\begin{equation}
\left( \calJ_\mu \right)_{mn}^{pq} \coloneqq \left\{
\begin{array}{l@{\;}l}
 1 & \text{if } \Delta E_{m' n'} = \Delta E_{p' q'}, \\
 \displaystyle\frac{\mu \eta}{i (\Delta E_{m' n'} - \Delta E_{p' q'})} & \text{otherwise.}
\label{eq:def_super_j}
\end{array}\right.
\end{equation}
We can now replace the resolvent $( i \left( \Delta E_{m' n'} - \Delta E_{p' q'} \right) + \mu \eta )^{-1}$ in Eq.~(\ref{eq:connection_r_pgp_coordinates}) by $(\mathcal{J}_\mu)_{mn}^{pq} / (\mu \eta)$ without changing anything in the limit $\eta \rightarrow 0^{+}$.

To obtain an index-free notation, we define the Schur product of two superoperators $\calA$ and $\mathcal{B}$ as
\begin{equation}
\left( \calA \circ \mathcal{B} \right)_{mn}^{pq} \coloneqq \left( \calA \right)_{mn}^{pq}  \left( \mathcal{B} \right)_{mn}^{pq},
\end{equation}
which allows us to reexpress Eq.~(\ref{eq:connection_r_pgp_coordinates}) as
\begin{equation}
\calP G^{(\mu)} \calP = \frac{1}{\mu \eta}\, \calJ_\mu \circ R^{(\mu)}.
\end{equation}
With this, the TCL generator given by Eq.~(\ref{eq:orders_tcl_explicit}) reads as
\begin{eqnarray}
S &=& \sum_{j=0}^\infty\: (-1)^j \sum_{\mu_0, \dots, \mu_j = 1}^\infty R^{(\mu_0)} \Bigg( \calJ_{\mu_1} \circ \frac{R^{(\mu_1)}}{\mu_1 \eta} \Bigg) \nonumber \\
    && \times  \Bigg( \calJ_{\mu_2} \circ \frac{R^{(\mu_2)}}{\mu_2 \eta} \Bigg) \cdots \Bigg( \calJ_{\mu_j} \circ \frac{R^{(\mu_j)}}{\mu_j \eta} \Bigg).
\label{eq:s_expressed_with_r}
\end{eqnarray}
This quantity appears to diverge in the limit $\eta \rightarrow 0^{+}$ because of the explicit negative powers of $\eta$. Moreover, every order $R^{(\mu)}$ starting from the fourth also diverges for $\eta \rightarrow 0^+$. This problem is well known for the special case of the \textit{T}-matrix rate equations \cite{Turek02, Koch04, Timm08, Begemann10, Koller10, Timm11} but persists in the generalized \textit{T}-matrix generator with coherences. We will show that all the singularities cancel, leaving all orders of the TCL generator well defined.

\subsection{Regularized \textit{T}-matrix generator and time-nonlocal memory kernel}

In preparation for the proof that all singularities in the TCL generator cancel, we first introduce a regularized version of the generalized \textit{T}-matrix generator $R(t,t_0)$ defined in Eq.\ (\ref{eq:r_definition}). As noted above, the expansion terms $R^{(\mu)}$ in Eq.\ (\ref{eq:orders_r}), starting from $\mu=4$, diverge for $\eta \rightarrow 0^{+}$. 

We first show that a systematic regularization can be achieved by inserting $\calQ$ projections after every $\calL_V$ (except the last) in Eq.~(\ref{eq:orders_r}). This is formally achieved by replacing $G$ in the last expression in Eq.\ (\ref{eq:r_definition}) by $G_{\calQ}$. Specifically, we define a regularized \textit{T}-matrix generator, or regularized \textit{T}-matrix for short, as
\begin{equation}
\Rreg(t,t_0) \coloneqq -i \calP \calL(t) G_{\calQ}(t,t_0) \calP,
\label{eq:def_r_reg}
\end{equation}
where the orders $\Rreg^{(\mu)}$ in the limit $t_0\to-\infty$ are given by
\begin{eqnarray}
\lefteqn{\Rreg^{(\mu)} \ketbra{p}{q} = -i\, e^{\mu \eta t} e^{i \calL_S t}\, \calP \calL_V \calQ}
  \nonumber \\
&&{}\times \frac{1}{\Delta E_{p'q'} - \calL_0 + (\mu-1) i \eta}\, \calL_V \calQ \cdots
  \calL_V \calQ \nonumber \\
&&{}\times \frac{1}{\Delta E_{p'q'} - \calL_0 + i \eta}\,
  \calL_V \calP e^{-i \calL_S t} \calP \ketbra{p}{q}.
\label{eq:orders_rreg}
\end{eqnarray}
Compared to Eq.\ (\ref{eq:orders_r}), $\calQ$ has been inserted between each pair of $\calL_V$ superoperators. Loosely speaking, this prevents the bath from returning to the reference state $\rho_B$ before the final (leftmost) $\calL_V$ acts.

The finiteness of the regularized \textit{T}-matrix is shown by a detour to the time-nonlocal ME: Here, one writes the evolution of the projected density matrix, in the interaction picture, as
\begin{equation}
\frac{\partial}{\partial t}\, \calP \rho(t) = \int_{t_0}^t ds\, \Sigma_K(t,s)\, \calP \rho(s),
\end{equation}
where $\Sigma_K$ is the time-nonlocal memory kernel. The explicit form of $\Sigma_k$ can be derived using projection superoperators (see, for example, Ref.\ \cite{BreuerPetruccione02}), with the result
\begin{equation}
\Sigma_K(t,s) = - \calP \calL(t) G_\calQ(t,s) \calQ \calL(s) \calP,
\label{eq:def_time_nonlocal_kernel}
\end{equation}
where we have used that $\calQ \rho(t_0) = 0$ and $\calP \calL \calP = 0$. The advantage of the time-nonlocal approach is that the kernel $\Sigma_K$ is always well defined, i.e., it does not rely on the existence of certain inverses, in contrast to the TCL generator. Furtermore, the expansion of $\Sigma_K$ is known to be well behaved and can, in principle, be generated to arbitrary order, for example using diagrammatic techniques~\cite{Schoeller94, Koenig95, Koenig96}.

Next, we show that $\Rreg$ can be expressed in terms of the zero-frequency limit of the Laplace transform of the time-nonlocal memory kernel $\Sigma_K$. From Eq.~(\ref{eq:def_time_nonlocal_kernel}) together with Eq.~(\ref{eq:deriv_propagator_lower_GQ}), we see that
\begin{equation}
\Sigma_K(t,s) = - \frac{\partial}{\partial s}\, \Rreg(t,s).
\end{equation}
This implies
\begin{equation}
\Rreg(t,t_0) = \Rreg(t,t) + \int_{t_0}^t ds\, \Sigma_K(t,s)
\label{eq:r_reg_integral_sigma_k}
\end{equation}
and thus for $t_0\to -\infty$,
\begin{equation}
\Rreg(t,-\infty) = -i \calP \calL(t) \calP + \int_{-\infty}^t ds\, \Sigma_K(t,s).
\end{equation}
On the right-hand side, the first term remains finite in the limit $\eta\to 0^+$ and is zero for the specific coupling $H_V$. The regularized \textit{T}-matrix thus exists for $\eta\to 0^+$ if the integral in the second term converges in this limit. The limit corresponds to the Hamiltonian $\Htot' = H_S + H_B + H_V$, which lacks the exponential time dependence. The memory kernel $\Sigma'_K(t-s) \coloneqq \lim_{\eta\to 0+} \Sigma_K(t,s)$ then only depends on the time difference since the Hamiltonian is time independent. In the resulting expression
\begin{equation}
\lim_{\eta\to 0^+} \Rreg(t,-\infty)
  = \int_0^\infty d\tau\, \Sigma'_K(\tau),
\label{eq:r_reg_integral_sigma_k_2}
\end{equation}
the integral represents the zero-frequency limit of the Laplace transform of the memory kernel of a system without the factor $e^{\eta t}$, i.e., of a model in which we do not turn on the hybridization slowly.

Moreover, the time-nonlocal memory kernel $\Sigma_K$ or $\Sigma'_K$ is always well defined for reservoirs with continuous spectrum. This is a consequence of the projections $\calQ$ in Eq.\ (\ref{eq:def_time_nonlocal_kernel}) that remove reducible contributions, which are responsible for the divergences \cite{Timm08, Schoeller09, Koller10}. The zero-frequency Laplace transform also exists, except in pathological cases: the integral over $\Sigma'_K$ in Eq.\ (\ref{eq:r_reg_integral_sigma_k_2}) only diverges if the memory kernel has at least one eigenvalue that decays with time $\tau$ like $1/\tau$ or even more slowly. For a small system with a finite number of relevant degrees of freedom, this cannot happen unless some degrees of freedom are completely decoupled. Therefore, $\Rreg$ and all its expansion terms remain finite in the limit $\eta \rightarrow 0^+$ for a generic system.

\subsection{Explicit \texorpdfstring{$\eta$}{n} dependence of the TCL generator}

The problem in exploiting Eq.~(\ref{eq:s_expressed_with_r}) is that the $R^{(\mu)}$ diverge for $\mu \geq 4$, in addition to the explicit divergences due to the $\eta$ in the denominators. In order to take the limit $\eta \rightarrow 0^{+}$, we first make every dependence on $\eta$ explicit. We start by writing out all the singular contributions to the $R^{(\mu)}$. To do so, we define
\begin{eqnarray}
\lefteqn{\Rreg^{(\mu, \mu')} \ketbra{p}{q}
    \coloneqq -i\, e^{(\mu-\mu')\eta t} e^{i \calL_S t}\, \calP \calL_V \calQ}
    \nonumber \\
&&{}\times \frac{1}{\Delta E_{p'q'} - \calL_0 + (\mu-1) i \eta}\, \calL_V \calQ \nonumber \\
&&{}\times \frac{1}{\Delta E_{p'q'} - \calL_0 + (\mu-2) i \eta} \cdots \calL_V \calQ
  \nonumber \\
&&{}\times \frac{1}{\Delta E_{p'q'} - \calL_0 + (\mu' + 1) i \eta}\, \calL_V
     \calP e^{-i \calL_S t} \ketbra{p}{q}.\hspace{1.8em}
\label{eq:def_rreg_parts}
\end{eqnarray}
The only difference compared to $\Rreg^{(\mu-\mu')}$ given in Eq.\ (\ref{eq:orders_rreg}) lies in different prefactors in front of the $i \eta$ in the denominators. We note that $\Rreg^{(\mu)}=\Rreg^{(\mu, 0)}$. Now we can make all the singular parts of $R^{(\mu)}$ explicit by inserting unit operators $\mathcal{I}=\calP+\calQ$ after every $\calL_V$ in Eq.~(\ref{eq:orders_r}) and by using the superoperators $\calJ_\mu$ defined in Eq.~(\ref{eq:def_super_j}):
\begin{eqnarray}
R^{(\mu)} &=& \sum_{m=0}^{\mu-1}\; \sum_{\mu > \mu_1 > \dots > \mu_m > 0}
        \Rreg^{(\mu,\mu_1)} \Bigg[ \calJ_{\mu_1} \circ \Bigg( \frac{\Rreg^{(\mu_1,\mu_2)}}{\mu_1 \eta} \Bigg[ \calJ_{\mu_2} \nonumber \\
&&{} \circ \cdots \frac{\Rreg^{(\mu_{m-1},\mu_m)}}{\mu_{m-1} \eta} \Bigg( \calJ_{\mu_m} \circ
\frac{\Rreg^{(\mu_m,0)}}{\mu_m \eta} \Bigg) \Bigg] \Bigg),
\label{eq:orders_r_through_rreg_1}
\end{eqnarray}
where the $m=0$ term is understood as $\Rreg^{(\mu,0)}$.

Equation (\ref{eq:orders_r_through_rreg_1}) contains products that are alternating between common matrix products and Schur products. To make the structure more explicit and simplify the notation, we define a $\star$-product between an even number of superoperators $\calA_1, \dots, \calA_{2 n}$ by
\begin{eqnarray}
\calA_1 \star \dots \star \calA_{2 n} &\coloneqq& \calA_1 \circ \Big( \calA_2 \Big[ \calA_3 \circ \Big( \calA_4 \cdots
  \nonumber \\
&&{}\times \Big[ \calA_{2 n-1} \circ \calA_{2 n} \Big] \Big) \Big] \Big).
\end{eqnarray}
The $\star$-product is linear since both the common superoperator product and the Schur product are linear:
\begin{eqnarray}
\lefteqn{\calA_1 \star \cdots \star \left( c\mathcal{B}_1 + \mathcal{B}_2 \right) \star \cdots = c \calA_1 \star \cdots \star \mathcal{B}_1 \star \cdots} \nonumber \\
&&{} + \calA_1 \star \cdots \star \mathcal{B}_2 \star \cdots ,\hspace{12em}
\end{eqnarray}
where $c$ is any complex number. We can now reexpress Eq.~(\ref{eq:orders_r_through_rreg_1}) as
\begin{eqnarray}
R^{(\mu)} &=& \sum_{m=0}^{\mu-1}\; \sum_{\mu > \mu_1 > \dots > \mu_m > 0} \Rreg^{(\mu,\mu_1)} \nonumber \\
&&{}\times \left( \calJ_{\mu_1} \star \frac{\Rreg^{(\mu_1,\mu_2)}}{\mu_1 \eta} \star \cdots \star \calJ_{\mu_m} \star \frac{\Rreg^{(\mu_m,0)}}{\mu_m \eta} \right).\hspace{1.5em}
\end{eqnarray}
Inserting the last equation into Eq.~(\ref{eq:s_expressed_with_r}) makes all the singular parts of the TCL generator explicit:
\begin{widetext}
\begin{eqnarray}
S &=& \sum_{j=0}^\infty\;
    \sum_{m_0,\dots,m_j=0}^\infty\, \sum_{\substack{\mu_{00}, \dots, \mu_{j m_j} > 0 \\ \mu_{k 0} > \cdots > \mu_{k m_k}}}
    (-1)^j\, \Rreg^{(\mu_{00}, \mu_{01})} \left( \calJ_{\mu_{01}} \star \frac{\Rreg^{(\mu_{01}, \mu_{02})}}{\mu_{01} \eta} \star \calJ_{\mu_{02}} \star \frac{\Rreg^{(\mu_{02}, \mu_{03})}}{\mu_{02} \eta} \star \cdots \star \calJ_{\mu_{0 m_0}} \star \frac{\Rreg^{(\mu_{0 m_0}, 0)}}{\mu_{0 m_0} \eta} \right) \nonumber \\
&&{}\times \left( \calJ_{\mu_{10}} \star \frac{\Rreg^{(\mu_{10}, \mu_{11})}}{\mu_{10} \eta} \star \calJ_{\mu_{11}} \star \cdots \star \frac{\Rreg^{(\mu_{1 m_1}, 0)}}{\mu_{1 m_1} \eta} \right) \cdots \left( \calJ_{\mu_{j 0}} \star \frac{\Rreg^{(\mu_{j 0}, \mu_{j 1})}}{\mu_{j 0} \eta} \star \calJ_{\mu_{j 1}} \star \cdots \star \frac{\Rreg^{(\mu_{j m_j}, 0)}}{\mu_{j m_j} \eta} \right).
\label{eq:tcl_with_star_product_1}
\end{eqnarray}
It will prove advantageous to replace the summation variables $\mu_{ij}$ pairwise through $(\mu_{00},\mu_{01}) \rightarrow (\mu_0,\mu_0')$, $(\mu_{01},\mu_{02}) \rightarrow (\mu_1,\mu_1')$, \dots, $(\mu_{j m_j},0) \rightarrow (\mu_p,\mu_p')$. For this to be correct, the vector $u=(\mu_0,\mu_0', \dots, \mu_p, \mu_p')$ has to be an element of the set
\begin{equation}
U_p \coloneqq \left \{ (\mu_0,\mu_0', \dots, \mu_p, \mu_p') \in \mathbb{N}_0^{2(p+1)}
 \middle| (\mu_i > \mu_i') \ \land \ (\mu_i' = \mu_{i+1} \lor \mu_i'=0) \ \land \ \mu_p'=0 \right \}.
\end{equation}
For every such vector $u \in U_p$, we define the index set $Z(u) \coloneqq \{ i \in \mathbb{N}_0 | \mu_i' = 0 \}$, which contains the indices of the $\mu_i'$ that are zero. The summation variable $j$ in Eq.\ (\ref{eq:tcl_with_star_product_1}) is then given by $j=|Z(u)|-1$ and we can write the sum as 
\begin{eqnarray}
S &=& \sum_{p=0}^\infty\, \sum_{u \in U_p} (-1)^{|Z(u)|-1}\, \frac{\eta^{-p}}{\mu_1 \mu_2 \cdots \mu_p}\, \Rreg^{(\mu_0, \mu_0')}
    \Big( \calJ_{\mu_1} \star \Rreg^{(\mu_1, \mu_1')} \star \dots \star \calJ_{\mu_{n_1}} \star \Rreg^{(\mu_{n_1},0)} \Big)  \nonumber \\
&&{}\times \Big( \calJ_{\mu_{n_1 + 1}} \star \Rreg^{(\mu_{n_1 + 1}, \mu_{n_1 + 1}')} \star \dots \star \calJ_{\mu_{n_1+n_2}} \star \Rreg^{(\mu_{n_1+n_2},0)} \Big) \cdots
    \Big( \calJ_{\mu_{n_1 + \cdots + n_{j-1} + 1}} \star \dots \star \calJ_{\mu_p} \star \Rreg^{(\mu_{p},0)} \Big).
\label{eq:tcl_with_star_product_2}
\end{eqnarray}
\end{widetext}
We note that we always have $n_1+\cdots+n_j=p \in Z(u)$ because $\mu_p'=0$. Furthermore, since the $p=0$ term is  equal to $\Rreg$ we can interpret the TCL generator as the regularized generalized \textit{T}-matrix plus corrections~$\mathcal{K}$:
\begin{equation}
S = \Rreg + \calK.
\label{eq:s_as_t_mat_and_corrections}
\end{equation}
While $\Rreg$ is not singular, all terms for $p\ge 1$ in Eq.~(\ref{eq:tcl_with_star_product_2}), i.e., the corrections $\mathcal{K}$, diverge as $\eta^{-p}$. To prove that the limit $\eta \rightarrow 0^{+}$ is well defined, we thus have to expand all the $\Rreg^{(\mu, \nu)}$ and $\calJ_{\mu}$ in $\eta$, and show that every coefficient in front of a term with a negative power of $\eta$ is zero. Every term with a positive power of $\eta$ vanishes for $\eta \rightarrow 0^+$, and only the terms with $\eta^0$ will remain as the finite limit.

We start this program by inserting the Taylor expansion of $\calJ_\mu$ in $\eta$, which is given by
\begin{equation}
\calJ_\mu = \calJ^{(1)} + \mu \eta \calJ^{(0)}, 
\end{equation}
where
\begin{eqnarray}
\big( \calJ^{(0)} \big)_{mn}^{pq} &\coloneqq& \left\{
\begin{array}{ll@{}}
 0 & \text{if } \Delta E_{m' n'} = \Delta E_{p' q'}, \\
 \displaystyle\frac{-i}{\Delta E_{m' n'} - \Delta E_{p' q'}} & \text{otherwise,}
\end{array}\right. 
\label{eq:def_calJ_0} \\
\big( \calJ^{(1)} \big)_{mn}^{pq} &\coloneqq& \left\{
\begin{array}{ll}
 1 & \text{if } \Delta E_{m' n'} = \Delta E_{p' q'}, \\[0.5ex]
 0 & \text{otherwise.}
\end{array}\right.
\label{eq:def_calJ_1}
\end{eqnarray}
The $\star$-product is not associative in the sense that
\begin{equation}
\calA_1 \star \cdots \star  \calA_{2 n} \neq \left(\calA_1 \star \cdots \star \calA_{2 j}\right) \left(\calA_{2 j + 1} \star \cdots \star \calA_{2 n} \right).
\end{equation}
In Appendix \ref{app:proof_star_associativity}, we show, however, that $\star$-products of the following special form are associative:
\begin{eqnarray}
\lefteqn{\calJ^{(\lambda_1)} \star \calA_1 \star \cdots \star \calJ^{(1)} \star \calA_k \star \cdots \star \calJ^{(\lambda_n)} \star \calA_n}
\nonumber \\
&&\quad = \big( \calJ^{(\lambda_1)} \star \cdots \star \calJ^{(\lambda_{k-1})} \star \calA_{k-1} \big) \nonumber \\
&&\qquad{}\times \big( \calJ^{(1)} \star \calA_k \star \cdots \star \calJ^{(\lambda_{n-1})} \star \calA_{n} \big).\hspace{2.5em}
\label{eq:star_associativity}
\end{eqnarray}
Of course, the right-hand side of this equation can be decomposed further, by applying it recursively whenever one of the $\lambda_i$ is unity. In doing so, we find that Eq.~(\ref{eq:star_associativity}) can be written in the form
\begin{eqnarray}
\lefteqn{\left( \calJ^{(\lambda_1)} \star \calA_1 \star \calJ^{(0)} \star \cdots \star \calJ^{(0)} \star \calA_i \right)} \nonumber \\
&&{} \times \left( \calJ^{(1)} \star \calA_{i+1} \star \calJ^{(0)} \star \cdots \star \calJ^{(0)} \star \calA_j \right) \cdots \nonumber \\
&&{} \times \left( \calJ^{(1)} \star \calA_{u+1} \star \calJ^{(0)} \star \cdots \star \calJ^{(0)} \star \calA_n \right)
\end{eqnarray}
(recall that $\lambda_i \in \{0,1\}$). By inserting the Taylor expansion of every $\calJ_\mu$ into Eq.~(\ref{eq:tcl_with_star_product_2}) and decomposing every $\star$-product as described, we can write the TCL generator as a (matrix) product of $r$ $\star$-products (with $1 \leq r \leq p$), where the $i$th $\star$-product consists of $2 n_i$ factors. Furthermore, instead of specifying the vector $u=(\mu_0,\mu_0', \dots, \mu_p, \mu_p')$ in Eq.~(\ref{eq:tcl_with_star_product_2}), we can equivalently specify the differences
\begin{equation}
m_i = \mu_i - \mu_i'
\label{eq:def_mi}
\end{equation}
together with the information which of the $\mu_i'$ are zero, i.e., with the index set $Z(u)$. We can fully determine $Z(u)$ by specifying the numbers $\rho_0, \ldots, \rho_r \in \{0,1 \}$ such that
\begin{equation}
\mu_{n_1+\cdots+n_j}' =
\left\{\begin{array}{ll}
        0 & \text{if } \rho_j = 1, \\[0.5ex]
        \mu_{n_1+\cdots+n_j}-m_{n_1+\cdots+n_j} & \text{otherwise.}
\label{eq:mu_prime_in_z}
\end{array}\right.
\end{equation}
Thus, we always have $\rho_r=1$. The quantities $\mu_0, \dots, \mu_p$ now have to be understood as functions of $n_1, \dots, n_r$, $\rho_0, \dots, \rho_{r-1}$, and $m_1, \dots, m_p$ that are recursively given by 
\begin{eqnarray}
\mu_0 &=& m_0 + (1-\rho_0)\, \mu_1,
\label{eq:mu_recursive_relation_0} \\
\mu_1 &=& m_1 + \cdots + m_{n_1} +  (1-\rho_1)\, \mu_{n_1+1},
\label{eq:mu_recursive_relation_1} \\
\mu_k &=& m_k + \cdots + m_{n_1 + \cdots + n_{j+1}} \nonumber \\
&&{}+ (1-\rho_{j+1})\, \mu_{n_1 + \cdots + n_{j+1}+1},
\end{eqnarray}
where the last equation holds for $n_1 + \cdots + n_{j} + 1 \le k \le n_1 + \cdots + n_{j+1}$.
Similarly, the $\mu_i'$ are understood as functions of the same independent variables specified by either Eq.\ (\ref{eq:def_mi}) or Eq.\ (\ref{eq:mu_prime_in_z}). This enables us to write the correction term $\calK$ in Eq.\ (\ref{eq:s_as_t_mat_and_corrections}) as
\begin{widetext}
\begin{eqnarray}
\calK &=& \sum_{p=1}^{\infty}\, \sum_{r=1}^{p}\, \sum_{n_1+\cdots+n_r=p}\, \sum_{m_0, \dots, m_p = 1}^\infty\, \sum_{\rho_0, \dots, \rho_{r-1}=0}^{1}\, \sum_{\lambda_1=0}^1\, \sum_{\lambda_2=1-\rho_1}^1 \cdots \sum_{\lambda_r=1-\rho_{r-1}}^1 \frac{(-1)^{\rho_0+\cdots+\rho_{r-1}}\, \eta^{-\Sigma_\lambda}}{\mu_1^{\lambda_1} \mu_{n_1+1}^{\lambda_2} \cdots \mu_{n_1+\cdots+n_{r-1}+1}^{\lambda_r}}\, \Rreg^{(\mu_0, \mu_0')} \nonumber \\
&&{}\times \Big( \calJ^{(\lambda_1)} \star \cdots \star \calJ^{(0)} \star \Rreg^{(\mu_{n_1}, \mu_{n_1}')} \Big)
\Big( \calJ^{(\lambda_2)} \star \cdots \star \calJ^{(0)} \star \Rreg^{(\mu_{n_1+n_2}, \mu_{n_1+n_2}')} \Big) \cdots
    \Big( \calJ^{(\lambda_{r})} \star \cdots \star \calJ^{(0)} \star \Rreg^{(\mu_p, 0)} \Big),\hspace{1.5em}
\label{eq:corrections_to_rreg_1}
\end{eqnarray}
with $\Sigma_\lambda \coloneqq \lambda_1 + \cdots + \lambda_p$. The final step is to insert the Taylor expansion in $\eta$ of all $\Rreg^{(\mu,\mu')}$, which is given by
\begin{equation}
\Rreg^{(\mu,\mu')} = -i \sum_{k_1, \dots, k_{m-1}=0}^\infty (-i \eta)^{k_1+\cdots + k_{m-1}}\, (\mu-1)^{k_1} \cdots (\mu'+1)^{k_{m-1}}\, \big[k_1, \dots, k_{m-1} \big],
\end{equation}
where $\mu-\mu'=m$ and we have defined the superoperator $[k_1, \dots, k_{m-1}]$ acting as
\begin{eqnarray}
[k_1, \dots, k_{m-1}] \ketbra{p}{q} &\coloneqq& \lim_{\eta\rightarrow 0} e^{i \calL_S t}\, \calP \calL_V \calQ  \left( \frac{1}{\Delta E_{p' q'} -\calL_0 +(\mu-1)i \eta} \right)^{\!k_1+1} \calL_V \calQ \left( \frac{1}{\Delta E_{p' q'} -\calL_0 +(\mu-2)i \eta} \right)^{\!k_2+1} \cdots \notag \\
&&{}\times \calL_V \calQ \left( \frac{1}{\Delta E_{p' q'} -\calL_0 +(\mu'+1)i \eta} \right)^{\!k_{m-1}+1} \calL_V \calP\, e^{-i \calL_S t} \ketbra{p}{q}.
\label{eq:def_bracket}
\end{eqnarray}
At first glance, this does not seem to be well defined, since the right-hand side depends on $\mu$ and $\mu'$ whereas the left-hand side does not. However, the superoperators $[k_1, \dots, k_{m-1}]$ do not depend on the prefactors of $i \eta$, as long as they are positive, as was shown in Appendix B of Ref.\ \cite{Timm11}. Inserting the Taylor expansion into Eq.~(\ref{eq:corrections_to_rreg_1}) then finally makes every contribution of $\eta$ explicit:
\begin{eqnarray}
\calK &=& \sum_{p=1}^{\infty}\, \sum_{r=1}^{p}\, \sum_{n_1+\cdots+n_r=p}\, \sum_{m_0, \dots, m_p = 1}^\infty\, \sum_{\lambda_1, \dots, \lambda_r=0}^1\; \sum_{k_{0,1 }, \dots, k_{0, m_0-1}, k_{1, 1}, \dots, k_{p, m_p-1} = 0}^\infty (-i)^{\Sigma_k+p+1}\, \eta^{\Sigma_k-\Sigma_\lambda}\, f_{p,r}(\bm{\lambda}, \bm{n}, \bm{m}, \vec{\bm{k}}) \nonumber \\
&&{}\times [\bm{k}_0]\, \big( \calJ^{(\lambda_1)} \star [\bm{k}_1] \star \calJ^{(0)} \star \cdots \star [\bm{k}_{n_1}] \big)\, \big( \calJ^{(\lambda_2)} \star [\bm{k}_{n_1+1}] \star \calJ^{(0)} \star \cdots \star [\bm{k}_{n_1+n_2}] \big) \cdots \big( \calJ^{(\lambda_r)} \star \cdots \star [\bm{k}_p] \big),
\label{eq:expansion_K_eta_explicit}
\end{eqnarray}
\end{widetext}
where we have used the shorthand notations
$\bm{\lambda} = (\lambda_1, \dots, \lambda_r)$,
$\bm{n} = (n_1, \dots, n_r)$,
$\bm{m} = (m_0, \dots, m_p)$,
$\bm{k}_i = (k_{i,1}, \dots, k_{i, m_{i-1}})$,
$\vec{\bm{k}} = (\bm{k}_0, \dots, \bm{k}_p)$, and
$\Sigma_k = \sum_{ij} k_{i,j}$.
The function $f_{p,r}$ reads as
\begin{eqnarray}
f_{p,r}(\bm{\lambda}, \bm{n}, \bm{m}, \vec{\bm{k}}) &\coloneqq& \sum_{\rho_0=0}^1\, \sum_{\rho_1=1-\lambda_2}^1 \cdots \sum_{\rho_{r-1}=1-\lambda_r}^1 g_0 g_1 \cdots g_p \nonumber \\
&&{}\times \frac{(-1)^{\rho_0+\cdots+\rho_{r-1}}}{\mu_1^{\lambda_1} \mu_{n_1+1}^{\lambda_2} \cdots \mu_{n_1+\cdots+n_{r-1}+1}^{\lambda_r}},
\label{eq:def_f_p_r}
\end{eqnarray}
with the definition
\begin{equation}
g_j \equiv g(\mu_j, \mu_j', \bm{k}_j) \coloneqq (\mu_j-1)^{k_{j,1}} \cdots (\mu_j'+1)^{k_{j,m_j-1}}.
\label{eq:def_gj}
\end{equation}
Recall that the $\mu_i$ and $\mu_i'$ are understood as functions of $\rho_0$, \dots, $\rho_{r-1}$, $\bm{n}$, and~$\bm{m}$.

It is not easy to give a physical interpretation of the terms in Eq.\ (\ref{eq:expansion_K_eta_explicit}). They all describe contributions to the time evolution of the projected density matrix $\calP\rho$, grouped into blocks of $\star$-products (in parentheses). Within each block, the small system does not return to any state with the same energy difference $\Delta E_{p' q'}$ as the initial state $\ketbra{p'}{q'}$, and so there is no dangerously divergent resolvent for $\eta \rightarrow 0^+$. Divergent factors of $1/\eta$ can, however, occur after (to the left of) each block: whenever the leftmost superoperator of a block is $\calJ^{(1)}$, the small system returns to a state with the same $\Delta E_{p' q'}$ [see Eq.\ (\ref{eq:def_calJ_1})], which is associated with a factor of $1/\eta$. This produces the factor $\eta^{-\Sigma_\lambda}$ in Eq.\ (\ref{eq:expansion_K_eta_explicit}) since $\Sigma_\lambda$ is the number of $\calJ^{(1)}$ that occur. The question whether $\calK$ is finite can now be decided by analyzing the prefactor function $f_{p,r}$: $\calK$ is finite if and only if $f_{p,r}$ vanishes whenever $\Sigma_k$ is less then~$\Sigma_\lambda$.

\subsection{Cancellation of divergences}

The detailed analysis of the function $f_{p,r}$ is relegated to Appendix \ref{app:analysis_prefactor_function}, as it is quite lengthy. Here, we present the result, which is remarkably simple. It turns out that $f_{p,r}$ is indeed zero if $\Sigma_k<\Sigma_\lambda$. The terms with $\Sigma_k>\Sigma_\lambda$ contain positive powers of $\eta$ and thus vanish for $\eta\to 0^+$. In the relevant case where $\Sigma_k=\Sigma_\lambda$, we find that, in spite of the complicated definition (\ref{eq:def_f_p_r}), $f_{p,r}$ only assumes the three possible values $-1$, $0$, $1$. To be specific, for every vector $\bm{k}_i$ denote the sum of its elements as $K_i = k_{i,1} + \cdots + k_{i, m_{i-1}}$. Now, consider the function $f_{p,r}(\bm{\lambda}, \bm{n}, \bm{m}, \vec{\bm{k}})$ for a fixed vector $\bm{\lambda}$ given by an arbitrary value of $\lambda_1$ and $\lambda_{j_1+1}=\lambda_{j_2+1}=\dots=\lambda_{j_{l-1}+1}=1$ for indices $0<j_1<\cdots<j_{l-1}<r$. Let every element of $\bm{\lambda}$ that we did not specify be zero, that is $\lambda_k=0$ if $k \notin \{ 1, j_1+1, \dots, j_{l-1}+1 \}$. Note that $l = \Sigma_\lambda - \lambda_1 +1$ and define a new vector $\bm{n}'$ with elements $n_k'=n_{j_{k-1}+1} + \cdots + n_{j_{k}}$. Then, $f_{p,r}$ is given by
\begin{eqnarray}
f_{p,r} &=& (-1)^{r-\Sigma_\lambda+\lambda_1-1}\, \Theta(K_0 \geq 1) \notag \\
&&{}\times \Theta(K_0+\cdots+K_{n_1'} \geq \lambda_1+1) \notag \\
&&{}\times \Theta(K_0+\cdots+K_{n_1'+n_2'} \geq \lambda_1+2)\cdots \notag \\
&&{}\times \Theta(K_0+\cdots+K_{n_1'+ \cdots + n_{l-1}'} = \lambda_1+l-1),\hspace{1.5em}
\label{eq:f_p_r_final}
\end{eqnarray}
where $\Theta(X)=1$ if $X$ is a true statement and $\Theta(X)=0$ otherwise. The $\Theta$ function in the last line of Eq.~(\ref{eq:f_p_r_final}) does not only imply that every possibly divergent term in the expansion of the TCL generator (where $\Sigma_k<\Sigma_\lambda$) is zero but it also guarantees that every $[\bm{k}_j]$ appearing to the right of the rightmost $\calJ^{(1)}$ has to be equal to $[\bm{0}]$. Furthermore, we see that the relevant quantity that organizes the terms is the vector $\bm{n}'$: its elements determine how many superoperators $[\bm{k}_j]$ appear between two $\calJ^{(1)}$.

Equation (\ref{eq:f_p_r_final}) shows that the numbers 
\begin{eqnarray}
M_0 &=& K_0,
\label{eq:M_K_1} \\
M_1 &=& K_1 + \cdots + K_{n_1'}, \\
M_j &=& K_{n_1'+ \cdots + n_{j-1}' + 1} + \cdots + K_{n_1'+ \cdots + n_{j}'},
\label{eq:M_K_3}
\end{eqnarray}
determine whether a certain term is allowed. In order to incorporate this additional structure into the expansion of the TCL generator, it is useful to reorganize the terms. To this end, we define
\begin{widetext}
\begin{eqnarray}
\lefteqn{[[M_0,M_1]]_\lambda \coloneqq \sum_{j=1}^\infty\, \sum_{n_1, \dots, n_j=1}^\infty
    \sum_{\substack{\bm{k}_0,\dots,\bm{k}_{n_1+\cdots+n_j} \\ K_0 = M_0,\: K_1 + \cdots + K_{n_1+\cdots+n_j} = M_1}}
    (-i)^{\lambda + n_1+\cdots+n_j + 1} (-1)^{j-1} [\bm{k}_0]} \nonumber \\
&&\quad{}\times \left( \calJ^{(\lambda)} \star [\bm{k}_1] \star \calJ^{(0)} \star \cdots \star [\bm{k}_{n_1}] \right) \left( \calJ^{(0)} \star [\bm{k}_{n_1+1}] \star \calJ^{(0)} \star \cdots \star [\bm{k}_{n_1+n_2}] \right) \cdots
    \left( \calJ^{(0)} \star \cdots \star [\bm{k}_{n_1+\cdots+n_j}] \right).
\end{eqnarray}
In this expression, the first superoperator $\calJ$ has an upper index of $\lambda$ while every other $\calJ$ has an upper index of zero. The vectors $\bm{k}_i$ come from the same set as in Eq.~(\ref{eq:expansion_K_eta_explicit}). In a similar fashion, we define
\begin{eqnarray}
[[M]] &\coloneqq& \sum_{j=1}^\infty\, \sum_{n_1, \dots, n_j=1}^\infty
    \sum_{\substack{\bm{k}_1,\dots,\bm{k}_{n_1+\cdots+n_j} \\ K_1 + \cdots + K_{n_1+\cdots+n_j} = M}}
    (-i)^{n_1+\cdots+n_j+1} (-1)^{j-1} \left( \calJ^{(1)} \star [\bm{k}_1] \star \calJ^{(0)} \star \cdots \star [\bm{k}_{n_1}] \right) \notag \\
&& {}\times \left( \calJ^{(0)} \star [\bm{k}_{n_1+1}] \star \calJ^{(0)} \star \cdots \star [\bm{k}_{n_1+n_2}] \right) \cdots
    \left( \calJ^{(0)} \star \cdots \star [\bm{k}_{n_1+\cdots+n_j}] \right),
\end{eqnarray}
where the first $\calJ$ has an upper index of unity while every other $\calJ$ has an upper index of zero. This means that every block (in parentheses) except for the leftmost returns the state of the small system to one with a different energy difference. On the other hand, the leftmost block in $[[M_0,M_1]]_1$ and $[[M]]$ always returns the state it acts on to one with the same energy difference [see Eqs.\ (\ref{eq:def_calJ_0}) and (\ref{eq:def_calJ_1})]. As noted above, this is exactly the case where a divergent factor $1/\eta$ appears. Note that the quantities $n_i$ in the last two equations correspond to the elements of the vector $\bm{n}'$ introduced earlier but we have dropped the primes in order to keep the notation as compact as possible. The corrections $\calK$ can now be written as
\begin{eqnarray}
\calK &=& \sum_{l=1}^\infty\, \sum_{\lambda=0}^1\, \sum_{M_0, \dots, M_{l} = 0}^\infty \Theta(M_0 \geq 1)\, \Theta(M_0+M_1 \geq \lambda+1) \cdots \Theta(M_0+\cdots+M_{l-1} = \lambda + l - 1) \nonumber \\
&& {}\times [[M_0, M_1]]_\lambda\: [[M_2]] \cdots [[M_l]] \nonumber \\
&=& \sum_{l=1}^\infty\, \sum_{\lambda=0}^1\, \sum_{M_0=1}^{\lambda+l-1}\,
    \sum_{M_1=\max(0,\lambda+1-M_0)}^{\lambda+l-1-M_0} \cdots \sum_{M_{l-2}=\max\big(0,\lambda+l-2-\sum_{i=0}^{l-3}M_i\big)}^{\lambda+l-1-\sum_{i=0}^{l-3} M_i} \nonumber \\
&&{}\times [[M_0, M_1]]_\lambda\: [[M_2]] \cdots [[M_{l-2}]]\: [[M_{l-1} = \lambda+l-1-\textstyle \sum_{i=0}^{l-2} M_i]]\: [[M_l = 0]].
\label{eq:expansion_K_double_brackets}
\end{eqnarray}
\end{widetext}
Here, $l$ is the number of double brackets and every $[[M_0,M_1]]_1$ and $[[M]]$ is associated with exactly one factor of $1/\eta$. In this sense, Eq.\ (\ref{eq:expansion_K_double_brackets}) expresses the correction $\calK$ in terms of the elementary singularities. $[[M_0,M_1]]_0$ is not associated with a negative power of $\eta$. On the other hand, the sum of all $M_j$ equals the sum of all $K_j$ [see Eqs.\ (\ref{eq:M_K_1})--(\ref{eq:M_K_3})], which equals $\Sigma_k$. This is the exponent of the product of all positive powers of $\eta$ resulting from the Taylor expansion of the regularized \textit{T}-matrix. The sums in Eq.\ (\ref{eq:expansion_K_double_brackets}) restrict $\Sigma_k$ to equal $\lambda+l-1$. But this is just the total negative power of $\eta$ from the divergences. Hence, Eq.\ (\ref{eq:expansion_K_double_brackets}) contains only the terms that survive for $\eta\to 0^+$.

At every order $l$, only a finite number of terms contribute to Eq.\ (\ref{eq:expansion_K_double_brackets}). We can therefore systematically expand the corrections $\calK$ using the introduced double brackets. The first few terms are given by
\begin{eqnarray}
\calK &=& \ [[1,0]]_1 + [[1, 0]]_0\: [[0]] + [[1, 1]]_1\: [[0]] + [[2, 0]]_1\: [[0]] \notag \\
&&{} + [[1, 0]]_0\: [[1]]\: [[0]] + [[1, 1]]_0\: [[0]]\: [[0]] + \cdots
\label{eq:first_correction_terms}
\end{eqnarray}
This is of course not an expansion in the bare coupling, since every double bracket contains infinitely many $\calL_V$. But given a system-bath coupling of the form of  Eq.~(\ref{eq:hamiltonian_v}), every $[[M_0,M_1]]_\lambda$ is at least of fourth order and every $[[M]]$ is at least of second order. Since $\calK$ stands for the correction term in Eq.\ (\ref{eq:s_as_t_mat_and_corrections}) one needs to add the regularized \textit{T}-matrix $\Rreg$ to obtain the TCL generator. We thus find that up to the second order, the TCL generator is equal to the regularized \textit{T}-matrix. Additionally, we see that the fourth-order correction has to come from the first term in Eq.~(\ref{eq:first_correction_terms}), and the sixth-order corrections from the terms in the first line, specifically
\begin{eqnarray}
\calK^{(4)} &=& [[1, 0]]_1^{(4)}, \\
\calK^{(6)} &=& [[1, 0]]_0^{(4)}\: [[0]]^{(2)} + [[1, 1]]_1^{(4)}\: [[0]]^{(2)} \notag \\
&& {}+ [[2, 0]]_1^{(4)}\: [[0]]^{(2)},
\end{eqnarray}
where the upper index indicates the order in $\calL_V$. Noting that the regularized \textit{T}-matrix is given by $\Rreg = -i [0] -i [0,0,0] - \cdots$, it is now straightforward to explicitly calculate the first orders of the TCL generator, where we reinsert the definitions of the double brackets:
\begin{eqnarray}
S^{(2)} &=& -i\, [0],
\label{eq:expansion_tcl_2} \\
S^{(4)} &=& -i\, [0,0,0] + i\, [1] \left( \calJ^{(1)} \star [0] \right), \\
S^{(6)} &=& -i\, [0,0,0,0,0] + [1] \left( \calJ^{(1)} \star [0] \star \calJ^{(0)} \star [0] \right) \notag \\
        && {}+ i\, [1] \left( \calJ^{(1)} \star [0,0,0] \right) \notag \\
        && {}- [1] \left( \calJ^{(1)} \star [0] \right) \left( \calJ^{(0)} \star [0] \right) \notag \\
        && {}+ i\, \big( [1,0,0] + [0,1,0] + [0,0,1] \big) \left( \calJ^{(1)} \star [0] \right) \notag \\
        && {}+ [1] \left( \calJ^{(0)} \star [0] \right) \left( \calJ^{(1)} \star [0] \right) \notag \\
        && {}-i\, [1] \left( \calJ^{(1)} \star [1] \right) \left( \calJ^{(1)} \star [0] \right) \notag \\
        && {}-i\, [2] \left( \calJ^{(1)} \star [0] \right) \left( \calJ^{(1)} \star [0] \right).
\label{eq:expansion_tcl_6}
\end{eqnarray}
We have obtained an expansion of the TCL generator in terms of superoperators that are finite by construction. This is the central result of this paper. Higher-order terms can of course be generated using appropriate computer algebra systems but become increasingly lengthy.

\section{Summary and Conclusions}

In this paper, we have shown how to systematically expand the TCL generator in the coupling between the relevant small system and the environment. By comparing the TCL generator and the propagator of the reduced density matrix order by order, we have derived a representation of the orders of the TCL generator in terms of the orders of the propagator [Eq.~(\ref{eq:orders_tcl_explicit_n})]. Although this expansion and the one in terms of ordered cumulants are equivalent, the former is expected to be advantageous for numerical calculations and for analytical resummation schemes because it requires fewer terms. This indicates that massive cancellations, which are not obvious a priori, must occur in the ordered cumulants. Additionally, we have shown that the same series can be rewritten as a recursion relation [Eq.~(\ref{eq:orders_tcl_recursive})], which further simplifies the expansion.

In order to use the TCL formalism to describe transport through quantum dots beyond weak dot-lead coupling, insight into higher-order terms is required. With this motivation, we have generalized the approach of Ref.\ \cite{Timm11}, in which all orders of the Pauli TCL generator (limited to diagonal elements of the reduced density matrix) were derived, to describe the complete reduced density matrix with coherences. As in the Pauli case, seemingly divergent terms ultimately inherited from the \textit{T}-matrix occur. However, we have shown that all divergent terms cancel order by order, leaving finite corrections to be added to the properly regularized \textit{T}-matrix to obtain the TCL generator. An interesting mathematical structure is revealed, in which the product alternates between the Schur (elementwise) product and the usual matrix product. We have derived a relatively compact form of the expansion to arbitrary order, which only contains finite terms [Eq.\ (\ref{eq:expansion_K_double_brackets})]. The fundamental objects appearing in this expansion describe processes that are each associated with an ``elementary'' singularity (a factor $1/\eta$, where $\eta$ is the adiabatic-switching rate sent to zero at the end of the calculations). It thus provides an expansion in the most singular contributions, not one in the dot-lead coupling. However, on its basis, the expansion of the TCL generator in the coupling can be obtained explicitly [Eqs.~(\ref{eq:expansion_tcl_2})--(\ref{eq:expansion_tcl_6})], where our results guarantee the finiteness of all terms.

Clearly, outside the perturbative regime, there is no reason for any term in the series to be small, and the convergence of the perturbative series is not guaranteed. The next step, made possible by the present work, will be to derive the conditions for its convergence. Moreover, since every term is formally known one might hope to perform partial resummations. Since the TCL ME is exact and local in time, not just the stationary state but the full dynamics in the intermediate-coupling regime can be explored.

\acknowledgments

The authors thank M. R. Wegewijs for useful discussions. K. N. acknowledges financial support by the Deut\-sche Forschungsgemeinschaft through Research Training Group RTG 1995.

\vspace{1ex}

\appendix

\section{\label{app:proof_identity_inverse}Proof of Eq.~(\ref{eq:simplification_inverse})}

We define $A \coloneqq \left[ \calP G + G_{\calQ} \calQ \right]^{-1} \calP$ and write $G = \mathcal{I} + \bar G$ and $G_{\calQ} = \mathcal{I} + \bar G_{\calQ}$, where $\bar G$ and $\bar G_{\calQ}$ are given by 
\begin{eqnarray}
\bar G(t,t_0) &=& -i \int_{t_0}^{t} ds\, \calL(s) G(s, t_0), \\
\bar G_{\calQ}(t,t_0) &=& -i \calQ \int_{t_0}^{t} ds\, \calL(s) G(s, t_0).
\end{eqnarray}
With these definitions, we find
\begin{equation}
\calP = \left( \mathcal{I} + \calP \bar G + \bar G_{\calQ} \calQ \right) A.
\label{eq:deriv_app_identity_inverse_1}
\end{equation}
By letting $\calQ$ act from the left on Eq.\ (\ref{eq:deriv_app_identity_inverse_1}) we obtain
\begin{equation}
0 = \left( \calQ + \bar G_{\calQ} \calQ \right) A = \left( \mathcal{I} + \bar G_{\calQ} \right) \calQ A = G_{\calQ} \calQ A
\label{eq:deriv_app_identity_inverse_2}
\end{equation}
since $\calP \calQ=0$ and $\calQ \bar G_{\calQ} = \bar G_{\calQ}$. It is guaranteed that the inverse of $G_{\calQ}$ exits, and by letting it act from the left on Eq.~(\ref{eq:deriv_app_identity_inverse_2}) we see that $\calQ A = 0$, or equivalently $A = \calP A$. Thus, by replacing $A \rightarrow \calP A$ in Eq.~(\ref{eq:deriv_app_identity_inverse_1}), the term $\bar G_{\calQ} \calQ$ vanishes, and after taking the inverse of the remaining factor to the left of $A$, we find
\begin{equation}
A = \calP \left[ \mathcal{I} + \calP \bar G \calP \right]^{-1} = \calP \left[ \calQ + \calP G \calP \right]^{-1},
\end{equation}
which completes the proof.

\section{\label{app:proof_recursive_expansion}Recursive expansion of the TCL generator}

In this appendix, we prove Eq.\ (\ref{eq:orders_tcl_recursive}). We start from Eq.\ (\ref{eq:orders_tcl_explicit_n}) and proceed by induction. The proof for the case $n=1$ is obvious. To prove the induction step, we require the identity
\begin{eqnarray}
\lefteqn{\sum_{\mu_1 + \dots + \mu_{j} = \mu} f_j(\mu_1,\dots,\mu_j)} \nonumber \\
&& \quad= \sum_{k=1}^{\mu-j+1} \sum_{\mu_1 + \dots + \mu_{j-1} = \mu - k} f_j(\mu_1,\dots,\mu_{j-1},k), \label{eq:summenregel_2}
\end{eqnarray}
where $f_j$ is an arbitrary function of $j$ arguments and $\mu_1, \dots, \mu_j$ are greater then zero. We also need that
\begin{equation}
\sum_{j=2}^\mu\, \sum_{k=1}^{\mu-j+1} \cdots = \sum_{k=1}^{\mu-1}\, \sum_{j=2}^{\mu-k+1} \cdots
\label{eq:summenregel_3}
\end{equation}
Equipped with these equations, we now express the $n$th order of the TCL generator as 
\begin{widetext}
\begin{eqnarray}
S^{(n)} &=& \sum_{k=1}^n\, (-1)^{k+1}
    \sum_{n_1 + \dots + n_k = n}
    \calP \dot G^{(n_1)} \calP G^{(n_2)} \calP \cdots \calP G^{(n_k)} \calP \nonumber \\
    &=&  \calP \dot G^{(n)} \calP
            + \sum_{k=2}^n (-1)^{k+1} \sum_{n_1 + \dots + n_k = n} \calP \dot G^{(n_1)} \calP G^{(n_2)} \calP \cdots \calP G^{(n_k)} \calP \nonumber \\
    &\stackeq{eq:summenregel_2}&
        \calP \dot G^{(n)} \calP + \sum_{k=2}^n\, \sum_{j=1}^{n-k+1} (-1)^{k+1}
\sum_{n_1 + \dots + n_{k-1} = n -j} \calP \dot G^{(n_1)} \calP G^{(n_2)} \calP \cdots \calP G^{(n_{k-1})} \calP  G^{(j)} \calP \nonumber \\
    &\stackeq{eq:summenregel_3}&
        \calP \dot G^{(n)} \calP + \sum_{j=1}^{n-1}\, \sum_{k=2}^{n-j+1} (-1)^{k+1}
  \sum_{n_1 + \dots + n_{k-1} = n -j} \calP \dot G^{(n_1)} \calP G^{(n_2)} \calP \cdots \calP G^{(n_{k-1})} \calP  G^{(j)} \calP \nonumber \\
    &=& \calP \dot G^{(n)} \calP - \sum_{j=1}^{n-1} \left( \sum_{k'=1}^{n-j} (-1)^{k'+1}
  \sum_{n_1 + \dots + n_{k'}  = n -j} \calP \dot G^{(n_1)} \calP G^{(n_2)} \calP \cdots \calP G^{(n_{k'})} \calP \right) G^{(j)} \calP \nonumber \\
  &=& \calP \dot G^{(n)} \calP - \sum_{j=1}^{n-1} S^{(n-j)} \calP  G^{(j)} \calP.
\end{eqnarray}
\end{widetext}
In the second to last step we have introduced $k'=k-1$ and in the last step we have used the induction hypothesis. This proves the proposed recursion relation.

\section{\label{app:proof_star_associativity}Proof of Eq.~(\ref{eq:star_associativity})}

Consider the matrix elements of the left-hand side of Eq.~(\ref{eq:star_associativity}): 
\begin{eqnarray}
\lefteqn{\big( \calJ^{(\lambda_1)} \star \calA_1 \star \cdots \star \calJ^{(1)} \star \calA_k \star \cdots \star \calJ^{(\lambda_n)} \star \calA_n \big)_{mn}^{pq}} \nonumber \\
&& = \big( \calJ^{(\lambda_1)} \big)_{mn}^{pq}
\big( \calJ^{(\lambda_2)} \big)_{\alpha_1 \beta_1}^{pq} \cdots
\big( \calJ^{(1)} \big)_{\alpha_{k-1} \beta_{k-1}}^{pq} \cdots \nonumber \\
&& {}\times \big( \calJ^{(\lambda_n)} \big)_{\alpha_{n-1} \beta_{n-1}}^{pq} \big( \mathcal{A}_1 \big)_{mn}^{\alpha_1 \beta_1} \big( \mathcal{A}_2 \big)_{\alpha_1 \beta_1}^{\alpha_2 \beta_2} \cdots \big( \mathcal{A}_n \big)_{\alpha_{n-1} \beta_{n-1}}^{pq}. \nonumber \\
&& 
\label{eq:deriv_star_associativity}
\end{eqnarray}
Because $\big( \calJ^{(1)} \big)_{mn}^{pq}= \delta_{\Delta E_{m'n'}, \Delta E_{p'q'}}$ and the matrix elements of all $\calJ^{(\lambda)}$ only depend on differences of energies of states but not on the states themselves we have 
\begin{equation}
\big( \calJ^{(\lambda)} \big)_{\alpha \beta}^{pq}\, \big( \calJ^{(1)} \big)_{\gamma \delta}^{pq} = \big( \calJ^{(\lambda)} \big)_{\alpha \beta}^{\gamma \delta}\, \big( \calJ^{(1)} \big)_{\gamma \delta}^{pq}.
\end{equation}
Thus, with the substitutions $\alpha_{k-1} \rightarrow \alpha$ and $\beta_{k-1} \rightarrow \beta$, the right-hand side of Eq.~(\ref{eq:deriv_star_associativity}) can be written as 
\begin{equation}
\big( \calJ^{(\lambda_1)} \star \mathcal{A}_1 \star \cdots \star \mathcal{A}_{k-1} \big)^{\alpha \beta}_{m n}\,
\big( \calJ^{(1)} \star \mathcal{A}_k \star \cdots \star \mathcal{A}_{n} \big)_{\alpha \beta}^{p q},
\end{equation}
which completes the proof.

\section{\label{app:analysis_prefactor_function}Analysis of the prefactor function \texorpdfstring{$f_{p,r}$}{fpr}}

As a first step, we establish a connection between $f_{p,r}$, which is defined in Eq.\ (\ref{eq:def_f_p_r}), and the simpler function
\begin{equation}
\tilde f_{p,r}(\lambda_1, \bm{n}, \bm{m}, \vec{\bm{k}}) \coloneqq f_{p,r}\big( (\lambda_1, \underbrace{1, \dots, 1}_{\text{$r-1$ times}}), \bm{n}, \bm{m}, \vec{\bm{k}} \big).
\end{equation}
To do this, consider a fixed but arbitrary vector $\bm{\lambda}$ of dimension $r$, where $\lambda_1 \in \{0,1\}$, $\lambda_{j_1+1} = \lambda_{j_2+1} = \cdots = \lambda_{j_{l-1}+1} = 1$, and all other elements are zero, i.e., $k \notin \{1, j_1+1, j_2+1, \dots, j_{l-1}+1\} \Rightarrow \lambda_k=0$. Since the $j_i$ can always be chosen accordingly for any vector $\bm{\lambda}$ there is no loss of generality. Note that $l=\Sigma_\lambda-\lambda_1+1$. It then follows that $\rho_0,\rho_{j_1}, \dots, \rho_{j_{l-1}} \in \{0, 1\}$ and all other $\rho_k$ are equal to unity. Hence, we find
\begin{eqnarray}
&f_{p,r}&(\bm{\lambda}, \bm{n}, \bm{m}, \vec{\bm{k}}) \notag \\
    &=&\!\! \sum_{\rho_0, \rho_{j_1}, \dots, \rho_{j_{l-1}}=0}^1
    \frac{ (-1)^{r-l} (-1)^{\rho_0+\rho_{j_1}+\cdots+\rho_{j_{l-1}}} g_0 \cdots g_p }
    {\mu_1^{\lambda_1} \mu_{n_1+\cdots+n_{j_1}+1} \cdots \mu_{n_1+\cdots+n_{j_{l-1}}+1}} \notag \\
    &=& (-1)^{r-l} \sum_{\rho_0,\dots,\rho_{l-1}=0}^1 \frac{(-1)^{\rho_0+\cdots+\rho_{l-1}}g_0\cdots g_p}{\mu_1^{\lambda_1} \mu_{n_1'+1} \cdots \mu_{n_1'+\cdots+n_{l-1}'+1}} \notag \\
    &=& (-1)^{r-l}\, \tilde f_{p,l}(\lambda_1, \bm{n}', \bm{m}, \vec{\bm{k}}),
\label{eq:connection_f_and_tilde_f}
\end{eqnarray}
where we have defined the vector $\bm{n}'$ with elements $n_k'=n_{j_{k-1}+1} + \cdots + n_{j_{k}}$ for $k<l$ and $n_l'=p-n_1' - \cdots - n_{l-1}'$. Thus, in order to understand the behavior of $f_{p,r}$, it is sufficient to consider ~$\tilde f_{p,l}$.

Now for every vector $\bm{k}_j$ let $K_j = k_{j,1} + \cdots + k_{j, m_{j-1}}$ denote the sum of its elements. The definition (\ref{eq:def_gj}) of $g_j$ shows that $K_j=0$ implies $g_j=1$ and, in particular, that $K_0=0$ implies $g_0=1$. In this case, no part of the summand depends on $\rho_0$ except for the factor $(-1)^{\rho_0}$. Thus, if $K_0=0$ then $\tilde f_{p,l}=0$.

In the next step, we use the following fact: if $g(x)$ is a normalized polynomial in $x$ of order $n$, then $\sum_{\rho=0}^1 (-1)^\rho\, g(m+(1-\rho)\mu)$ is a normalized polynomial in $\mu$ of order $n$ without an order-zero term. With this, we can perform the sum over $\rho_0$ since $\mu_0 = m_0+(1-\rho_0)\mu_1$ [see Eq.\ (\ref{eq:mu_recursive_relation_0})] and we can interpret $g_0(\mu_0)=g_0(m_0+(1-\rho_0)\mu_1)$ as a polynomial of order $K_0$ in $\mu_1$ [see Eq.\ (\ref{eq:def_gj})]. We find
\begin{eqnarray}
\tilde f_{p,l} &=& \Theta(K_0 \geq 1) \sum_{\rho_1, \dots, \rho_{l-1}=0}^1 \,
    \frac{(-1)^{\rho_1+\cdots+\rho_{l-1}}\, g_1 \cdots g_p}{\mu_{n_1'+1}\cdots \mu_{n_1' + \cdots + n_{l-1}' + 1}} \notag \\
&& {}\times (\mu_1^{K_0-\lambda_1} + \cdots + c_1\, \mu_1^{1-\lambda_1}),
\label{eq:f_after_sum_rho0}
\end{eqnarray}
where $\Theta(X)=1$ if $X$ is a true statement and $\Theta(X)=0$ otherwise. The second line represents some polynomial in $\mu_1$ of order $K_0-\lambda_1$, with unit coefficient of the highest-order term and vanishing order-zero term. In this expression, only the factor $(-1)^{\rho_1}$, the functions $g_1, \dots, g_{n_1'}$, and the polynomial in the second line depend on $\rho_1$ (through $\mu_1, \dots, \mu_{n_1'}$). With the same reasoning as before, we conclude that $\tilde f_{p,r}$ is zero if $K_0+\cdots+K_{n_1'} = \lambda_1$ since in this case nothing except for the factor $(-1)^{\rho_1}$  depends on $\rho_1$. Therefore, $\tilde f_{p,l}$ is proportional to $\Theta(K_0+\cdots+K_{n_1'} \geq \lambda_1 + 1)$ and if this $\Theta$ function equals unity we can again interpret the second line of Eq.~(\ref{eq:f_after_sum_rho0}) together with the factor $g_1 \cdots g_{n_1'}$ as a polynomial in $\mu_1=m_1+\cdots+m_{n_1'}+(1-\rho_1)\mu_{n_1'+1}$ and perform the sum over $\rho_1$ to obtain a polynomial in $\mu_{n_1'+1}$ of order $K_0+\cdots+K_{n_1'+n_2'}-\lambda_1-1$.

The described procedure can now be iterated to eliminate all the remaining sums, leading to the result
\begin{eqnarray}
\tilde f_{p,l} &=& \ \Theta(K_0 \geq 1)\, \Theta(K_0+\cdots+K_{n_1'} \geq \lambda_1+1) \notag \\
&& {}\times \Theta(K_0+\cdots+K_{n_1'+n_2'} \geq \lambda_1+2) \cdots \notag \\
&& {}\times \Theta(K_0+\cdots+K_{n_1'+ \cdots + n_{l-1}'} \geq \lambda_1+l-1) \notag \\
&& {}\times g_{n_1' + \cdots + n_{l-1}' + 1} \cdots g_p.
\label{eq:f_p_l_after_sums}
\end{eqnarray}
Inspecting the last $\Theta$ function in Eq.~(\ref{eq:f_p_l_after_sums}) and noting that $l-1=\Sigma_\lambda-\lambda_1$, we see that a necessary condition for $\tilde f_{p,l}$ to be non-zero is that
\begin{equation}
\Sigma_k \geq K_0+\cdots+K_{n_1'+ \cdots + n_{l-1}'} \geq \lambda_1+l-1 = \Sigma_\lambda.
\end{equation}
Hence, we conclude that every term in the expansion of the TCL generator with a negative power of $\eta$ is zero [see Eq.~(\ref{eq:expansion_K_eta_explicit})], and therefore that the expansion is well behaved. Furthermore, in the limit $\eta \rightarrow 0^+$, only the terms with $\Sigma_k=\Sigma_\lambda$ remain. In this case, the last $\Theta$ function in Eq.~(\ref{eq:f_p_l_after_sums}) also implies that
\begin{equation}
K_{n_1'+\cdots+n_{l-1}'+1} = \cdots = K_p = 0,
\end{equation}
from which it follows that
\begin{equation}
g_{n_1' + \cdots + n_{l-1}' + 1} = \cdots =  g_p = 1.
\end{equation}
Substituting our result for $\tilde f_{p,l}$ into Eq.~(\ref{eq:connection_f_and_tilde_f}), we arrive at the result
\begin{eqnarray}
f_{p,r} &=& (-1)^{r-\Sigma_\lambda+\lambda_1-1}\, \Theta(K_0 \geq 1) \notag \\
&& {}\times \Theta(K_0+\cdots+K_{n_1'} \geq \lambda_1+1) \notag \\
&& {}\times \Theta(K_0+\cdots+K_{n_1'+n_2'} \geq \lambda_1+2)\cdots \notag \\
&& {}\times \Theta(K_0+\cdots+K_{n_1'+ \cdots + n_{l-1}'} = \Sigma_\lambda),
\end{eqnarray}
which was asserted in the main text.


\begin{thebibliography}{99}
\bibitem{footnote_RE}Equation (\ref{eq:G.0}) taken together with $\calQ \rho(t_0)=0$ implies that $\partial_t \calP \rho(t) = R(t,t_0)\, \calP \rho(t_0)$. Note the different argument of the last factor.
\end{thebibliography}
\end{document}